\documentclass[pra,preprint, twocolumn, subscriptaddress, showpacs,10pt]{revtex4}
\usepackage{amsmath}
\usepackage{graphicx}
\usepackage{hyperref}
\usepackage{ragged2e}

\begin{document}


\title{Measurement of the branching fractions and lifetime of the $5D_{5/2}$ level of Ba$^{+}$}



\author{Carolyn Auchter}
\thanks{Authors contributed equally}
\author{Thomas W. Noel}
\email[]{Corresponding author: noelt@u.washington.edu}
\author{\hspace{-1mm}$^{*}$\hspace{1mm}Matthew R. Hoffman}
\author{Spencer R. Williams}
\author{Boris B. Blinov}
\affiliation{Department of Physics, University of Washington, Seattle, Washington 98195, USA}
\newcommand{\Ba}{\textsuperscript{138}Ba\textsuperscript{+}}


\date{\today}

\begin{abstract}
We present a measurement of the branching fractions for decay from the long-lived $5D_{5/2}$ level in \Ba. The branching fraction for decay into the $6S_{1/2}$ ground state was found to be $0.846(25)_{\mathrm{stat}}(4)_{\mathrm{sys}}$. We also report an improved measurement of the $5D_{5/2}$ lifetime, $\tau_{5D_{5/2}}=31.2(0.9)$~s.  Together these measurements provide the first experimental determination of transition rates for decay out of the $5D_{5/2}$ level.  The low ($<7 \times 10^{-12}$~Torr) pressure in the ion trap in which these measurements were made simplified data acquisition and analysis.  Comparison of the experimental results with theoretical predictions of the transition rates shows good agreement.
\end{abstract}



\maketitle
\section{Introduction}
\paragraph*{}  Trapped ions provide a useful physical system for a variety of experimental applications including quantum computation \cite{Monroe14, Benhelm08}, precision measurement \cite{Orzel12, Hanneke08}, and frequency standards \cite{Rosenband08}.  These applications leverage the environmental isolation and long trap lifetimes provided by trapped ions to obtain long interrogation times and relatively systematic-error free methodologies.  A deep and thorough understanding of atomic theory is necessary for the development and analysis of trapped ion and other atomic physics experiments.  This understanding is cemented by the comparison of precisely measured atomic parameters with those calculated from atomic theory.  Such comparisons provide checks on the accuracy of the atomic wavefunctions involved in theoretical calculations.  Excited state lifetimes and branching fractions are readily measured and allow direct comparison with calculation results \cite{Gerritsma08, Ramm13, Nilsson10, Kreuter05}.  Here we present a measurement of the branching fractions for decay from the metastable $5D_{5/2}$ level in Ba$^{+}$. To our knowledge, a branching fraction measurement on such a long-lived atomic level as barium's $5D_{5/2}$ level ($\tau \sim$ 30 s) has not previously been done, although the branching fraction of the metastable $^{2} \hspace{-0.5 mm}D_{5/2}$ level in Yb$^{+}$ ($\tau \sim$ 7 ms) has been measured \cite{Taylor97}. The long lifetime of the metastable state in barium introduces additional systematic effects that must be carefully considered and controlled.  A discussion of these systematic effects follows the presentation of the results later in this article.

\section{Experimental Procedure}

\begin{figure}[th!!!]
\includegraphics[width=0.8\linewidth]{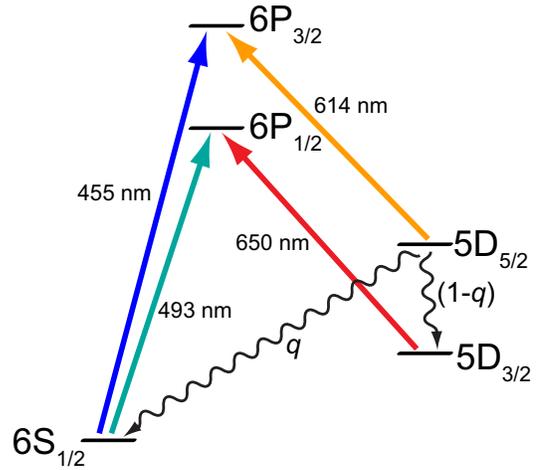}
\caption{(Color online) Relevant energy levels of \Ba.  Doppler cooling is done on the $6S_{1/2}$ to $6P_{1/2}$ transition with 493~nm light while 650~nm light is present to repump the $5D_{3/2}$ level. A laser at 455~nm allows us to shelve the ion in the long-lived $5D_{5/2}$ level by driving transitions between the $6S_{1/2}$ and $6P_{3/2}$ levels.  A shelved ion can be returned to the cooling cycle by applying light at 614~nm.  The two decay paths from the $5D_{5/2}$ level are shown.  The measurement of the branching fraction to the ground state, $q$, from that level is the subject of this article.}
\label{energy}
\end{figure}

\paragraph*{}  A single \Ba\ ion was confined in a linear Paul trap similar to that described in \cite{Gulde03, Noel14}. The combined action of a 20~l/s ion pump and periodic flashes of a titanium filament in a titanium sublimation pump maintained the pressure in the vacuum chamber at a pressure below the $7 \times 10^{-12}$~Torr minimum readable pressure of the Varian senTorr BA2C gauge controller with a Varian UHV-24p ionization gauge installed in the chamber.  The relevant level structure and transitions of a \Ba\ ion are displayed in Fig.~\ref{energy}. The ion was Doppler cooled primarily on the $6S_{1/2}$ to $6P_{1/2}$ transition with 493~nm light derived by frequency doubling a home-built 986~nm external cavity diode laser (ECDL) similar to the design described in \cite{Ricci95}. A second home-built ECDL at 650~nm was used to repump from the long-lived $5D_{3/2}$ level ($\tau \approx 80$~s \cite{Yu97}) to which the $6P_{1/2}$ level decays roughly one quarter of the time. In order to excite the ion to the $5D_{5/2}$ level, a home-built 455~nm ECDL was used to drive the $6S_{1/2}$ to $6P_{3/2}$ transition.  Since the $5D_{5/2}$ level is long-lived and disjoint from the cooling cycle, an ion occupying that level is said to be ``shelved," and the level itself is sometimes referred to as the ``shelved state." The $6P_{3/2}$ level branches upon decay into the $5D_{5/2}$, $5D_{3/2}$, and $6S_{1/2}$ levels with probabilities 0.215, 0.029, and 0.756, respectively \cite{Kurz08}. This means that in the presence of 455~nm and 650~nm light the ion will be reliably optically pumped into the shelved state.  Note that, in the presence of 455~nm light alone, an ion initially in the ground state will be shelved about 89\% of the time, with the remaining population transferred to the $5D_{3/2}$ level. Light at 614~nm, derived by frequency-doubling a 1228~nm ECDL, was used to deshelve the ion, when necessary, by driving the $5D_{5/2}$ to $6P_{3/2}$ transition, which returns the ion to the cooling cycle. Mechanical shutters were used on all lasers to ensure that only the desired wavelengths of light were interacting with the ion at any given time during the experimental cycle. 

\begin{figure}[t]
\includegraphics[width=\linewidth]{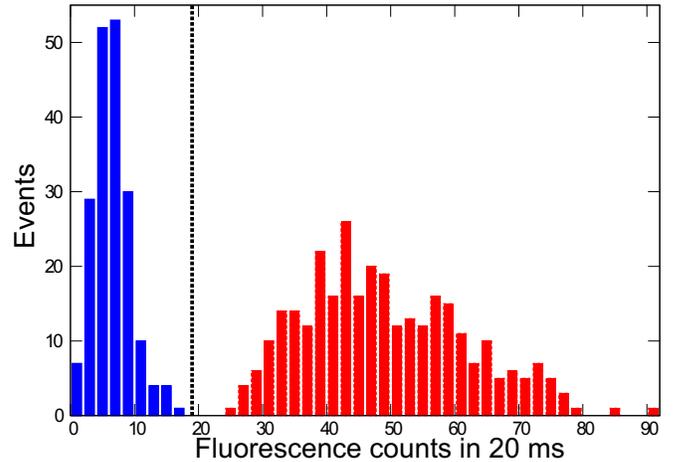}
\caption{(Color online) Histogram of the fluorescence counts recorded in the 500 runs comprising the 30~s wait time data point of the $5D_{5/2}$ lifetime measurement to be described later in this report.  The histogram is presented here as an example demonstrating the high quality of the discrimination between the bright and dark ion states.  This particular data was chosen because the number of events in the bright and dark peaks are roughly equal, making it easy to interpret visually without need for rescaling or other manipulation. The vertical dashed line shows a threshold below (equal to or above) which the run is categorized as dark (bright). Since the background (from laser scatter, room light, and PMT dark counts) does not vary significantly, the dark peak fits well to a Poisson distribution.  Modeling the bright counts is complicated by drift in the cooling laser frequency over the long duration of the data acquisition.  This implies that the bright distribution draws from a sum of Poisson distributions with a variety of means. Independently of how this fitting was done in detail, the probability for a false dark event (ion in the cooling cycle measured to have fluorescence below threshold) was invariably found to be $<10^{-3}$ and the probability for a false bright event was even lower.}
\label{histo}
\end{figure}

\begin{figure}[t]
\includegraphics[width=\linewidth]{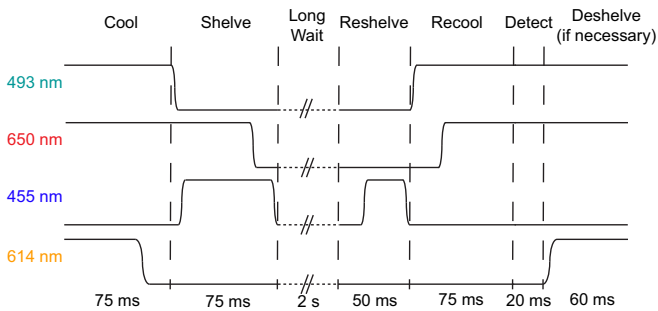}
\caption{(Color online) Timing sequence for application of each of the four lasers.  The ion is being exposed to a given laser when the corresponding trace in the diagram is high. The duration of application of each laser is largely determined by the slow response times of the mechanical shutters and the desire to have absolute confidence in the ordering of the sequence.  The operations involved are reliably completed in much shorter times than are used. The experimental cycle is detailed in the main text.  Briefly, the cycle consisted of Doppler cooling, shelving with 455~nm and 650~nm light, a 2~s wait for decays from the $5D_{5/2}$ level, re-shelving of the decays to ground state with 455~nm light, detection of the resulting ion fluorescence state, and deshelving when necessary.}
\label{waveforms}
\end{figure}

\paragraph*{}  The experimental cycle for the measurement of the branching fractions of the $5D_{5/2}$ level consisted of five major steps: cooling, shelving, wait for decay, re-shelving, and state detection.  First, the ion was Doppler cooled for 75~ms.  Next the ion was shelved to the $5D_{5/2}$ level with 65~ms of 455~nm light while the 650 nm light remained on to repump any decays to the $5D_{3/2}$ level. All light was then extinguished for a duration of 2~s, allowing the ion to decay in a few percent of the trials. Then the 455~nm beam was exposed for 30~ms in order to re-shelve any decay to the $6S_{1/2}$ ground state. Finally state detection was performed by turning on the cooling lasers, with an ion having decayed to the $5D_{3/2}$ level appearing ``bright" and an ion having decayed to the $6S_{1/2}$ ground state and been reshelved or not having decayed appearing ``dark." Fig.~\ref{histo} shows a histogram of fluorescence counts with clearly distinguished bright and dark distributions. Bright and dark state discrimination was $>99.9\%$. If the ion was found to be in the $5D_{5/2}$ level upon detection, then a deshelving pulse of 614~nm light was applied before restarting the cycle.  A waveform sequence detailing the timing of the application of each laser is shown in Fig.~\ref{waveforms}.

\section{Results}

\paragraph*{}  The branching fraction measurement comprised 13,000 experimental cycles like those discussed in the previous section. The probability that the ion was in the $5D_{5/2}$ level at the end of the procedure, $P_{\mathrm{dark}}$, depends on the initial shelving efficiency, $P_{\mathrm{sh}}$, the probability of decay from the $5D_{5/2}$ level during the wait time, $P_{\mathrm{dec}}$, the re-shelving efficiency, $P_{\textrm{re-sh}}$, the branching fraction for decay into the $6S_{1/2}$ level, $q$, and the probability of a sequential decay from $5D_{5/2}$ to $5D_{3/2}$ to $6S_{1/2}$ during the wait time, $P_{\textrm{seq-dec}}$. Following the experimental cycle, the ion will be found to be dark in three ways: (1) the ion was shelved and subsequently did not decay $P_{\mathrm{sh}}(1-P_{\mathrm{dec}})$, (2) the ion was shelved, subsequently decayed to the ground state, and was re-shelved $P_{\mathrm{sh}} P_{\mathrm{dec}} q P_{\textrm{re-sh}}$, and (3) the ion was shelved, subsequently decayed to the $5D_{3/2}$ level, subsequently decayed to the ground state, and was reshelved $P_{\mathrm{sh}} P_{\mathrm{dec}} (1-q) P_{\textrm{seq-dec}} P_{\textrm{re-sh}}$.  Adding these together, we find $P_{\mathrm{dark}}$ to be
\begin{equation}
\begin{array}{rcl}
P_{\mathrm{dark}} & = & P_{\mathrm{sh}}(1-P_{\mathrm{dec}})+P_{\mathrm{sh}}\; P_{\mathrm{dec}}\; q \; P_{\textrm{re-sh}}\vspace{1mm} \\ 
& &+P_{\mathrm{sh}} \; P_{\mathrm{dec}} (1-q) P_{\textrm{seq-dec}} \; P_{\textrm{re-sh}}.
\end{array}
\end{equation}
Solving this equation for $q$ shows how the branching fraction depends on all the quantities to be measured,
\begin{equation}
q=\frac{P_{\mathrm{dark}} - P_{\mathrm{sh}}\left[ 1 - P_{\mathrm{dec}} ( 1 - P_{\textrm{re-sh}} \; P_{\textrm{seq-dec}} ) \right]}{P_{\mathrm{sh}} \; P_{\mathrm{dec}} \; P_{\textrm{re-sh}}(1 -  P_{\textrm{seq-dec}})}.
\end{equation}
In order to extract the branching fractions from the data, we must independently measure each of these probabilities.  To measure $P_{\mathrm{sh}}$ and $P_{\textrm{re-sh}}$, we interleaved measurements of the shelving efficiency of an ion initially in the ground state both with the 650~nm laser unshuttered and shuttered with every 100 branching fraction cycles. Based on these measurements we observed no evidence of variation in the shelving efficiency which allowed us to consolidate all the data taken into a single set. These measurements yield the values of $P_{\mathrm{sh}}$ and $P_{\textrm{re-sh}}$ to be 0.999~49(42) and 0.888~7(16), respectively. We also independently measured the fraction of the time the ion decayed from the $5D_{5/2}$ level during the 2~s wait time by using the same procedure outlined in the previous section, but without the re-shelving pulse. This measurement finds $P_{\mathrm{dec}}=0.0654(22)$ and also provided the first data point in the $5D_{5/2}$ lifetime measurement discussed below.  The probability of the sequential decay, $P_{\textrm{seq-dec}}$, is small enough that an estimate based on the lifetime of the $5D_{3/2}$ level suffices to address the effect. We estimate that on average the decays to the $5D_{3/2}$ level reside there for one second before the fluorescence state is measured, which, using the level's 79.8~s lifetime \cite{Yu97}, implies that $P_{\textrm{seq-dec}}=0.012(6)$. We additionally correct for a slight bias toward ``bright" counts due to decays of the shelved ion that occur during the 75 ms of cooling before detection, resulting in an average of slightly less than three extra ``bright" counts per 1000 cycles. After this correction, we find the shelved fraction to be $P_{\mathrm{dark}}=0.9836(12)$. Putting all these measurements together, we find the branching fraction for decay of the $5D_{5/2}$ level to the $6S_{1/2}$ level to be $q=0.850(25)$ where the error is statistical, based on propagation of the binomial errors in all the measurements discussed above. 

\begin{figure}[tt!]
\includegraphics[width=\linewidth]{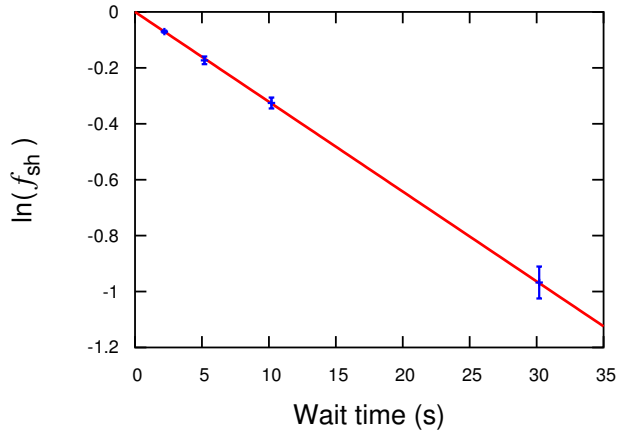}
\caption{(Color online) The natural logarithm of the fraction of trials in which the ion remained shelved, $f_{\mathrm{sh}}$, versus the duration that the ion spent in the $5D_{5/2}$ shelved state is plotted.  The errorbars on the four data points are statistical based on the binomial distribution.  The red line is a linear fit through the origin with slope its only fit parameter.  The fit implies the lifetime to be $\tau=31.1(0.9)$~s. }
\label{lifetime}
\end{figure}

\paragraph*{} The analysis above makes the unjustified assumption that in the $(1-P_{\mathrm{sh}})=0.05$\% of the cycles in which the ion failed to be shelved intially, it was always found to be bright upon fluorescence detection. It is reasonable to assume that whenever the ion was not shelved before the wait time, it must have resided either in the $5D_{3/2}$ level or the ground state, but we do not have a simple means of discerning which. If (when not initially shelved) the ion always resided in the $5D_{3/2}$ level at the onset of the wait time, then the analysis from the previous paragraph is valid, since the ion would indeed be found to be bright upon detection.  If, however, the ion always resided in the ground state as the wait time began, then the re-shelving pulse would have transferred the ion to the $5D_{5/2}$ level in 89\% of those trials.  Adjusting the analysis to assume that the ion was always in the ground state when the initial shelving pulse failed lowers the extracted value of the branching fraction to $q=0.842$(24).  The reality is likely that when the initial shelving pulse failed, some population resided in both the $5D_{3/2}$ and $6S_{1/2}$ levels. Therefore, including this effect as a systematic uncertainty, we report the measured value of the branching fraction for the decay of the $5D_{5/2}$ level into the $6S_{1/2}$ ground state to be $q=0.846(25)_{\mathrm{stat}}(4)_{\mathrm{sys}}$.

\paragraph*{} Both in order to extract transition rates from our branching fraction measurement and to help estimate systematic effects, we measured the lifetime of the $5D_{5/2}$ level by counting the number of decays with wait times of 2~s, 5~s, 10~s, and 30~s for 6000, 1000, 1000, and 500 experimental cycles respectively. The lifetime data is shown in Fig.~\ref{lifetime}, along with a line of best fit for the logarithm of the shelved fraction, $f_{\mathrm{sh}}$, versus the wait time. The only fit parameter is the lifetime $\tau=31.1(0.9)$~s where the error is statistical.

\paragraph*{}  In order to more directly compare our experimental result with theoretical predictions, we calculate transition rates from our measured lifetime and branching fraction for the decay from the $5D_{5/2}$ level into the $6S_{1/2}$ ground state. Using theoretical calculations for the relevant reduced matrix elements of the electric quadrupole (E2) and magnetic dipole (M1) operators, we find transition rates predicted by several theorists using the following equations:
\begin{equation}
\begin{split}
A^{E2}_{ki}&=\frac{1}{4\pi\epsilon_0}\frac{1}{15}\frac{\omega^5}{\hbar c^5}\frac{1}{g_k} \left| S_{ki} \right|^2 \\&\quad \text{where  } S_{ki}= \left|\left< k || \mathrm{E2} || i \right> \right| 
\label{E2}
\end{split}
\end{equation}
\begin{equation}
\begin{split}
A^{M1}_{ki}&=\frac{1}{4\pi\epsilon_0}\frac{4\omega^3}{3\hbar c^3}\frac{1}{g_k} \left| S_{ki} \right|^2 \\&\quad \text{where  } S_{ki}=\left| \left< k || \mathrm{M1} || i \right> \right|.
\label{M1}
\end{split}
\end{equation}
Here $\hbar \omega$ is the difference in energy between states $i$ and $k$, and $g_k$ is the degeneracy of the upper state, $k$ \cite{Sobelman72}. Our results and the transition rates predicted by theory are presented in Table \ref{theory}. While the results presented here are insufficiently precise to distinguish between the selected set of theoretical calculations, an increase in precision of about a factor of three on the transition rate from the $5D_{5/2}$ level to the $6S_{1/2}$ level would suffice to begin rejecting some calculations.

 \begin{table}[b!!]
 \caption{Comparison of theoretical predictions and our experimental results\label{theory}}
 \begin{ruledtabular}
 \begin{tabular}{cccc}
  Transition & \multicolumn{2}{c}{Present Work} & Previous Theory\\
 $\;$ & Relative Strength & $A_{ki} (\textrm{s}^{-1})$ & $A_{ki} (\textrm{s}^{-1})$\\
\hline
$5D_{5/2} \rightarrow 6S_{1/2}$ & $0.846(25)$ & $0.0271(15)$ & 0.02692\cite{Safronova11}\\ %
&&& 0.02693$^\ddagger$\cite{Guet07}\\ %
&&& 0.02607\cite{Gurell07}\\%
&&& 0.02800\cite{Sahoo06}\\%
$5D_{5/2} \rightarrow 5D_{3/2}$ & $0.154(25)$ & $0.00494(83)$ & 0.005470\cite{Safronova11}\\ %
&&& 0.005543\cite{Guet07}\\ %
&&& 0.005573\cite{Gurell07}\\%
&&& 0.005499$^\ddagger$\cite{Sahoo06}\\%
 \end{tabular}
 \end{ruledtabular}
\justify
\footnotesize $^\ddagger$ These values do not appear explicitly in the references, but were deduced based on the $5D_{5/2}$ lifetimes quoted therein.
\end{table}

\paragraph*{} The major systematic effects of experiments involving long dwell times in metastable states are effects related to the presence of background gas, leakage light, and ion heating. At our trap pressure of $<7 \times 10^{-12}$~Torr, the quenching effect due to the presence of background gas is nearly negligible.  Based on quenching rates for the $5D_{5/2}$ level found in literature \cite{Madej90, Yu97}, background gas collisions at our estimated trap pressure reduce the measured lifetime by 0.1~s.  Therefore we report the lifetime of the $5D_{5/2}$ level of \Ba\ to be $\tau=31.2(0.9)$~s. The use of mechanical shutters ensured that no unwanted laser light impinged on the ion during the long dwell time.  A number of mechanisms, not all of which are fully understood, contribute to heating of ions in Paul traps \cite{Turchette00}.  Thus it could be supposed that the long duration without cooling involved in each experimental cycle of the branching ratio and lifetime measurements could result in ions so hot that they would not reliably appear bright when the cooling lasers are finally exposed for detection. However, ion heating resulting in false dark readings was found to be a negligible effect. We ran 1000 cycles of an experimental sequence in which the ion spent the same duration in the absence of the cooling lasers as it did in the branching ratio measurement, but in the ground state instead of the $D_{5/2}$ level.  Ion heating is not expected to depend on electronic state, so the heating in this test should faithfully reproduce the heating in the branching ratio measurement.  After the duration spent in the dark, the cooling lasers were turned on and the ion was measured to be bright in every trial. Furthermore, looking at the distribution of the bright counts, we see no change in our ability to discriminate bright ions from dark ones due to heating during the long dwell time.

\section{Conclusions}

\paragraph*{} In conclusion, we have made the first experimental measurement of the branching fractions from the $5D_{5/2}$ level of Ba II, finding that the level decays to the ground state with probability 0.846$(25)_{\mathrm{stat}}(4)_{\mathrm{sys}}$. In addition, our measurement of the $5D_{5/2}$ lifetime, $\tau = 31.2(0.9)$~s, improves on the previous best measurement, $\tau = 32.3(2.6)$~s \cite{Yu97}, reducing the uncertainty by almost a factor of three. Together these measurements provide the first experimentally determined values for the rates of the $5D_{5/2}$ to $6S_{1/2}$ and $5D_{5/2}$ to $5D_{3/2}$ transitions.  The values for those transtion rates are found to be 0.0271(15)~s$^{-1}$ and 0.00494(83)~s$^{-1}$, respectively. Good agreement between our measured values and recent theoretical predictions was found.

\begin{acknowledgments}
The authors wish to thank John Wright, Chen-Kuan Chou, Tomasz Sakrejda, Anupriya Jayakumar, Richard Graham, and Zichao Zhou for helpful discussions. The authors would also like to acknowledge the work of Matt Bohman that contributed to this experiment. This research was supported by National Science Foundation  Grant No. 1067054.
\end{acknowledgments}

\bibliography{Epicbib}

\end{document}